\documentclass{article}

\usepackage{arxiv}
\usepackage[utf8]{inputenc}
\usepackage[T1]{fontenc}
\usepackage{hyperref}
\usepackage{float}
\usepackage{url}
\usepackage{booktabs}
\usepackage{placeins}
\usepackage{amsfonts}
\usepackage{nicefrac}
\usepackage{graphicx}
\usepackage{caption}
\usepackage{subcaption}
\usepackage[numbers]{natbib}
\usepackage{doi}
\hypersetup{hidelinks}

\title{SEGAA: A Unified Approach to Predicting Age, Gender, and Emotion in Speech}

\author{\href{https://orcid.org/0009-0003-8999-7272}{Aron R} \\
	School of Computer Science and Engineering\\
	Vellore Institute of Technology\\
	Chennai, India \\
	\texttt{aronritesh.2020@vitstudent.ac.in} \\
	\And
	\href{https://orcid.org/0009-0005-5788-119X}{Indra Sigicharla} \\
	School of Computer Science and Engineering\\
	Vellore Institute of Technology\\
	Chennai, India \\
	\texttt{indrakiran.sigicharla2020@vitstudent.ac.in} \\
	\And
	\href{https://orcid.org/0000-0003-3938-7495}{Dr K Mohanaprasad} \\
	School of Electronics Engineering\\
	Vellore Institute of Technology\\
	Chennai, India \\
	\texttt{kmohanaprasad@vit.ac.in} \\
	\And
	\href{https://orcid.org/0000-0001-7641-2149}{Sourabh Tiwari} \\
	Voice Intelligence Group\\
	Samsung R\&D Institute \\
	Bengaluru, India \\
	\texttt{sourabh.t@samsung.com} \\
	\And
	{Shivani Arora} \\
	Voice Intelligence Group\\
	Samsung R\&D Institute \\
	Bengaluru, India \\
	\texttt{shivani.ar@samsung.com} \\
	\And
	\href{https://orcid.org/0000-0001-9154-8669}{Dr Nithya Darisini P S} \\
	School of Electronics Engineering\\
	Vellore Institute of Technology\\
	Chennai, India \\
	\texttt{psnithyadarisini@vit.ac.in} \\
    \And
	{Chirag Periwal} \\
	School of Electronics Engineering\\
	Vellore Institute of Technology\\
	Chennai, India \\
	\texttt{chirag.periwal2020@vitstudent.ac.in} \\
}

\date{}

\renewcommand{\shorttitle}{\textit{Age, Gender, Emotion in Speech}}

\begin{document}
\maketitle

\begin{abstract}
	The interpretation of human voices holds importance across various applications. This study ventures into predicting age, gender, and emotion from vocal cues, a field with vast applications. Voice analysis tech advancements span domains, from improving customer interactions to enhancing healthcare and retail experiences. Discerning emotions aids mental health, while age and gender detection are vital in various contexts. Exploring deep learning models for these predictions involves comparing single, multi-output, and sequential models highlighted in this paper. Sourcing suitable data posed challenges, resulting in the amalgamation of the CREMA-D and EMO-DB datasets. Prior work showed promise in individual predictions, but limited research considered all three variables simultaneously. This paper identifies flaws in an individual model approach and advocates for our novel multi-output learning architecture Speech-based Emotion Gender and Age Analysis (SEGAA) model. The experiments suggest that Multi-output models perform comparably to individual models, efficiently capturing the intricate relationships between variables and speech inputs, all while achieving improved runtime.
\end{abstract}


\section{Introduction}
    In our increasingly digital world, the ability to glean profound insights from the nuances of human voices has assumed paramount importance. This paper delves into the captivating domain of predicting age, gender, and emotion based on vocal cues, a multidisciplinary field teeming with far-reaching applications.

     Voice analysis technologies have rapidly advanced, bringing transformative breakthroughs to various fields \citep{elena-bucea_assessing_2021}. These advancements not only optimize customer interactions but also hold the potential to revolutionize healthcare diagnostics, fundamentally altering our comprehension and engagement with human communication. In mental healthcare, the ability to discern emotions offers an opportunity to improve emotional and behavioral disorders \citep{jiao_behavioral_2020, mcteague_identification_2020}. Moreover, the application of this technology extends to the retail sector, where it enhances the consumer experience \citep{cachero-martinez_building_2021}. Simultaneously, the ability to detect age and gender variables finds relevance in evaluating the mental health requirements of distinct demographic groups \citep{gonzales_mental_2020}. It also plays a crucial role in e-services and policy formulation \citep{badircea_e-commerce_2021}. Thus, the ability to discern age, gender, and emotion from voice data emerges as a multifaceted field with a plethora of discernible applications. 

\subsection{Prior art}

    This research endeavour embarks on a comprehensive exploration of advanced deep learning architectures tailored for the prediction of age, gender, and emotional states. Additionally, it undertakes an exhaustive comparative analysis, meticulously examining diverse methodologies. These methodologies include individual models, where a single model is used to predict a single variable; multi-output models, which employ a single model to predict all three variables simultaneously; and sequential models, which cascade individual models to create a sequence representing the three variables at distinct stages. This scrutiny aims to elucidate the effectiveness of these approaches in the vital task of detecting these three pivotal variables.

    Before beginning the experiments, addressing the critical challenge of sourcing a dataset that encompasses all three target labels was necessary. Consequently, a rigorous review of popular, openly available speech datasets was undertaken to test the models. Datasets like RAVDESS and IEMOCAP \citep{busso_iemocap_2008, livingstone_ryerson_2018} were found to have emotion labels but lacked age and gender annotations. The TESS dataset \citep{dupuis_toronto_2010} featured only two speakers, rendering it unsuitable for our purposes despite including all three variables. Similarly, the DES dataset \citep{engberg_design_1997} comprised just four speakers, limiting its utility. The Common Voice dataset \citep{ardila_common_2020} had age and gender labels but omitted emotion labels. As a result, the CREMA-D and EMO-DB datasets \citep{burkhardt_database_2005, cao_crema-d_2014} were identified as the only suitable sources containing all three labels and adequate data. An aggregate dataset was created by amalgamating these two datasets to facilitate our experiments.

    Prior research has produced remarkable results in predicting individual variables using dedicated models. For instance, \citep{ahmed_ensemble_2023, tursunov_age_2021} achieved impressive accuracy rates of 90.47\% and 92.73\% on CREMA-D for emotion prediction using 1D CNNs and other deep learning architectures like GRU and LSTMs. Meanwhile, \citep{pappagari_copypaste_2021} attained an accuracy of 76.83\% using their copy-pasta architecture. Furthermore, \citep{mat_jizat_multilanguage_2021} detailed a multilingual speech-based gender classification method employing time-frequency features and the SVM classifier, achieving 81\% classification accuracy on the EMO-DB dataset for gender classification. Several studies \citep{goyal_gender_2020, tursunov_age_2021} have explored the prediction of age and gender from speech, employing diverse methodologies involving MLPs and CNNs. Notably, limited research has been conducted on predicting all three variables—namely, age, emotion, and gender—from speech data.

    While \citep{zaman_one_2021} proposed a "one-source-to-detect-all" solution for predicting age, gender, and emotion from one model, we identified methodological flaws in their approach. Specifically, these studies utilized three separate datasets to train three distinct tasks, a strategy that may not be optimal for enabling a model to discern all three variables from a single speech source.

    The motivation behind employing multi-output models lies in their ability to concurrently specify the relationships between multiple outcome variables (Y) and feature variables (X). As underscored by \citep{schmid_machine_2022}, this approach acknowledges the potential interdependencies between these outcome variables, resulting in enhanced statistical power or improved predictive accuracy. Moreover, \citep{schmid_machine_2022} observed that variations in dependency structures within the covariates influenced the performance of predictive methods. Notably, this impact was consistent for both univariate and multivariate approaches, indicating that neither approach held a definitive advantage concerning the dependency structures within the covariates.

    Given these considerations, this study initiates a series of experiments to determine whether multi-output models or univariate approaches more effectively classify age, gender, and emotion from speech. This paper aims to contribute valuable insights to the ongoing discourse surrounding the optimal approach for addressing this multifaceted predictive task. 

    The rest of the article is structured as follows: Section 2 details the methodology of various models used in this experiment, discusses the database and explains the feature extraction along with our novel SEGAA model. Section 3 elaborates on the results based on the experimental study. Section 4 presents the conclusion of this article. 

\section{Methodology}
\label{sec:method}

\subsection{Dataset Description}

    Given these considerations, this study initiates a series of experiments to determine whether multi-output models or univariate approaches more effectively classify age, gender, and emotion from speech. This paper aims to contribute valuable insights to the ongoing discourse surrounding the optimal approach for addressing this multifaceted predictive task.
    
    The EMO-DB database, a publicly accessible German emotional database, was developed by the Institute of Communication Science at the Technical University in Berlin, Germany. The data collection involved ten proficient speakers, equally distributed between genders (five males and five females). The database comprises 535 speech utterances covering seven distinct emotions: anger, boredom, anxiety, happiness, sadness, disgust, and neutrality.
   
    Along with the EMO-DB dataset, the CREMA-D dataset was also used. CREMA-D consists of 7,442 authentic speech clips from a diverse group of 91 actors. The actors include 48 males and 43 females, with ages ranging from 20 to 74. They represent diverse racial and ethnic backgrounds, including African American, Asian, Caucasian, Hispanic, and Unspecified. During the recording process, the actors delivered a set of 12 distinct sentences, each expressed with one of six specific emotions: Anger, Disgust, Fear, Happiness, Neutrality, and Sadness. The study focused on emotions, and data samples labelled "neutral" were integrated into the "Neutrality" class. 
   
    The primary goal with respect to the data augmentation part of this study is to enhance the diversity and quality of the dataset to improve the performance of emotion recognition models. The age and gender of the speakers were extracted from the speaker information provided and appended to the corresponding data sample. Consequently, our comprehensive dataset comprised a fusion of the EMO-DB and CREMA-D datasets. 
   
    The study focused on emotions, and data samples labelled "neutral" were integrated into the "Neutrality" class. Simultaneously, we omitted data samples with the label "boredom" as we narrowed our scope to include the following emotions: Anger, Disgust, Fear, Happiness, 
    Neutrality, and Sadness.

\begin{figure}[H]
    \centering
    \includegraphics[width=0.9\linewidth]{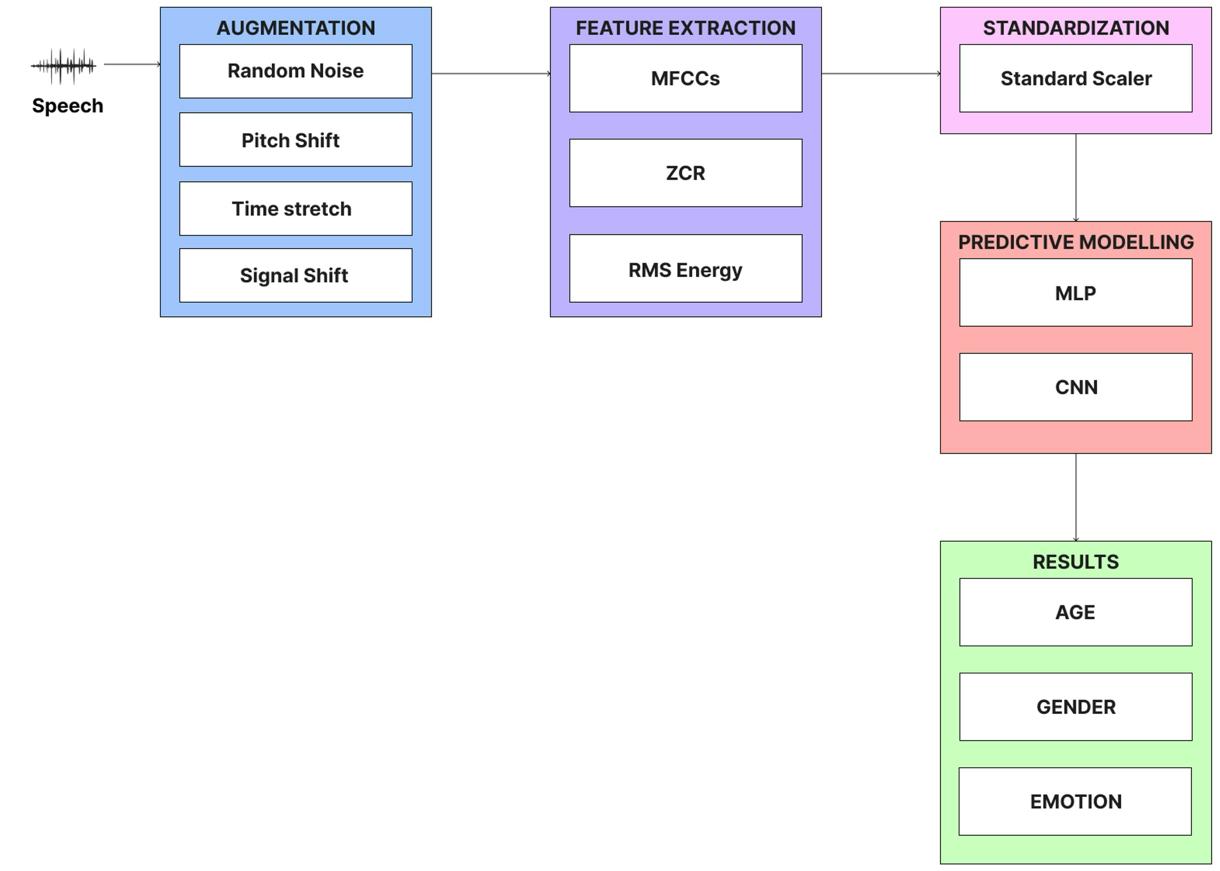}
    \caption{Workflow Diagram}
    \label{fig:workflow-diagram}
\end{figure}

\subsection{Feature Extraction / Pre-processing}

    We implemented audio processing functions to augment the original data samples. These functions include noise addition, time stretching, pitch shifting, and signal shifting. The noise addition process introduces random noise to the audio data, while time stretching alters the duration of the samples. Pitch shifting, on the other hand, modifies the audio pitch, and signal shifting shifts the audio signal along the time axis.
    
    Our feature extraction pipeline involves calculating essential audio features such as the Zero-Crossing Rate (ZCR), Root Mean Square Energy (RMSE), and Mel-frequency cepstral coefficients (MFCC). These features are widely used in audio and speech processing to capture relevant characteristics of the audio signals.
    
    The augmentation techniques were applied to the original audio samples to construct the final dataset. Audio features were extracted for each sample, including the original data and its augmented versions. This process enabled the generation of diverse features, encompassing pitch, duration, and signal properties variations. The resulting dataset showcases a significant increase in size and diversity.
    
    We split the dataset into input features (X) and target labels (Y). Y was one-hot encoded to represent the categorical variables. The dataset was then divided into training, testing, and validation sets, using a stratified approach to ensure class distribution balance. Input features were standardized using the Standard Scaler to enhance model convergence and performance. To maintain consistency, the testing and validation sets were transformed using the scaler fitted to the training set. This is then used to train and test the MLP and the SEGAA models, which finally output the age, gender and emotion results as described in Figure 1. The results of the respective models will then be compared and inferred from.

    Sigmoid activation is used for gender prediction with binary cross-entropy loss, while softmax activation is employed for emotion and age prediction with categorical cross-entropy. The optimization process relies on Stochastic Gradient Descent (SGD) with a learning rate of 0.0005, a decay rate of 1e-6, momentum set to 0.9, and Nesterov momentum. The model undergoes training for 200 epochs with a batch size of 32, during which validation accuracy metrics for gender, emotion, and age predictions are continuously monitored.

\subsection{Models}
\subsubsection{Individual Models}
\paragraph{Individual MLP:}

    The designed multi-layer perceptron architecture incorporates fully connected layers. It commences with an input layer customized to accommodate the extracted features, followed by several hidden layers. The hidden layers consist of 2048, 1024, 512, and 64 neurons as described in figure 2, equipped with Rectified Linear Unit (ReLU) activation functions. A dropout layer with a rate of 0.25 is introduced to address the overfitting issue. For making predictions specific to each label, distinct activation functions are applied in the output layers.

    Sigmoid activation is used for gender prediction with binary cross-entropy loss, while softmax activation is employed for emotion and age prediction with categorical cross-entropy. The optimization process relies on Stochastic Gradient Descent (SGD) with a learning rate of 0.0005, a decay rate of 1e-6, momentum set to 0.9, and Nesterov momentum. The model undergoes training for 200 epochs with a batch size of 32, during which validation accuracy metrics for gender, emotion, and age predictions are continuously monitored.
    
\begin{figure}[h]
    \centering
    \includegraphics[width=0.9\linewidth]{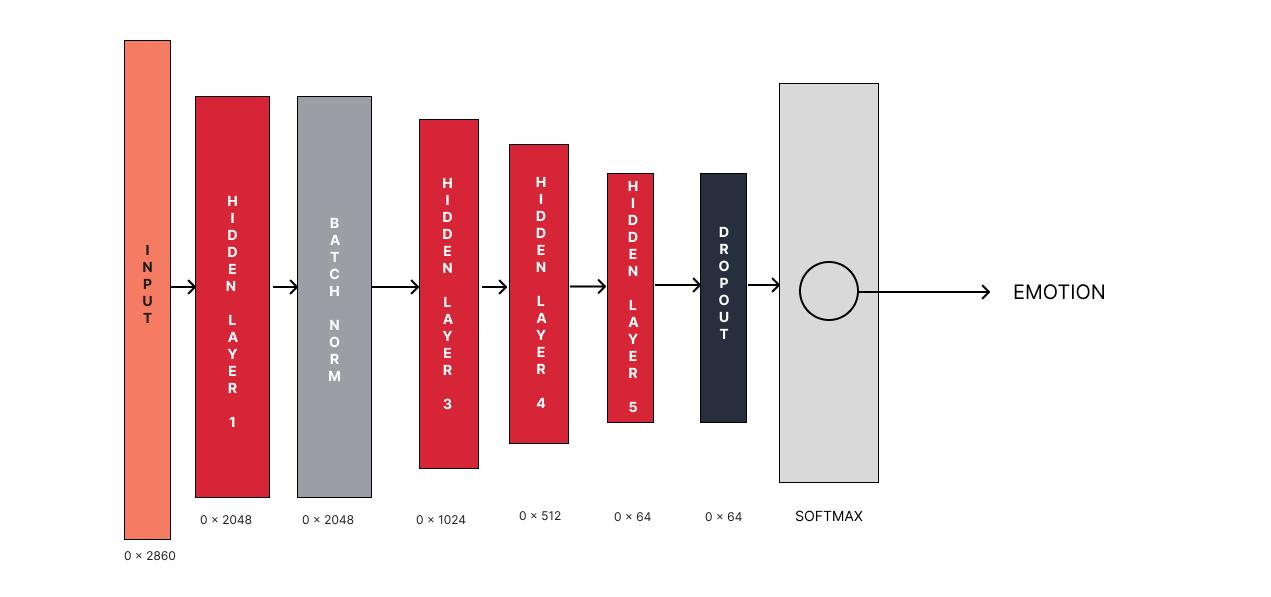}
    \caption{Model architecture for Individual MLP}
    \label{fig:individual-mlp}
\end{figure}

\paragraph{Individual SEGAA (Gen - 0):}

    The network comprises several key layers in this architecture. Initially, there is a convolutional layer featuring 256 filters, a kernel size of 5, and a stride of 1. This layer is augmented with batch normalization and is succeeded by max pooling with a pool size of 5 and a stride of 2. Following this, another convolutional layer has 128 filters, a kernel size of 5, and a stride of 1. This layer incorporates batch normalization, max pooling with a pool size of 5 and a stride of 2, and dropout at a rate of 20\%. Finally, the architecture culminates with a convolutional layer employing 64 filters, a kernel size of 5, and a stride of 1. Similar to the previous layers, it includes batch normalization and max pooling.
    
    After these convolutional layers, a flattening operation is performed, leading to a shared densely connected layer comprised of 32 neurons. Each of these neurons benefits from batch normalization and dropout, with a rate set at 20\%. The hyperparameters include a convolutional kernel size of 5, a dropout rate of 20\%, the utilization of the Adam optimizer for efficient optimization, and the implementation of an early stopping strategy with a patience of 5 epochs. A learning rate reduction strategy was also implemented, featuring a patience of 3 epochs and a reduction factor of 0.5.

    The models employed for 'emotion' and 'age' predictions utilize softmax activation functions in their final layers, each comprising six neurons. In contrast, the 'gender' prediction model employs binary softmax activation in its final layer, which consists of 2 neurons. 

\paragraph{Individual SEGAA:}

    This model architecture is an improvement made upon the previously mentioned SEGAA Gen-0. The architecture comprises three convolutional blocks, each followed by Batch Normalization, Max Pooling, and Dropout layers, facilitating feature extraction and dimensionality reduction. Subsequently, a Flatten layer consolidates the extracted information, leading to a densely connected layer with 64 neurons, further normalized and regularized using Batch Normalization and Dropout as described in figure 3. Key hyperparameters include a kernel size of 3 and a stride of 1 for the Convolutional layers, dropout rates set at 0.3, and the adoption of the Nadam optimizer for efficient optimization. The model undergoes training for 200 epochs with a batch size of 16, and validation accuracy metrics are continuously monitored. Early stopping and learning rate reduction callbacks are strategically used to ensure optimal convergence. 
    
    The models employed for 'emotion' and 'age' predictions utilize softmax activation functions in their final layers, comprising six neurons. In contrast, the 'gender' prediction model employs binary softmax activation in its final layer, which consists of 2 neurons. 

\begin{figure}[H]
    \centering
    \includegraphics[width=\linewidth]{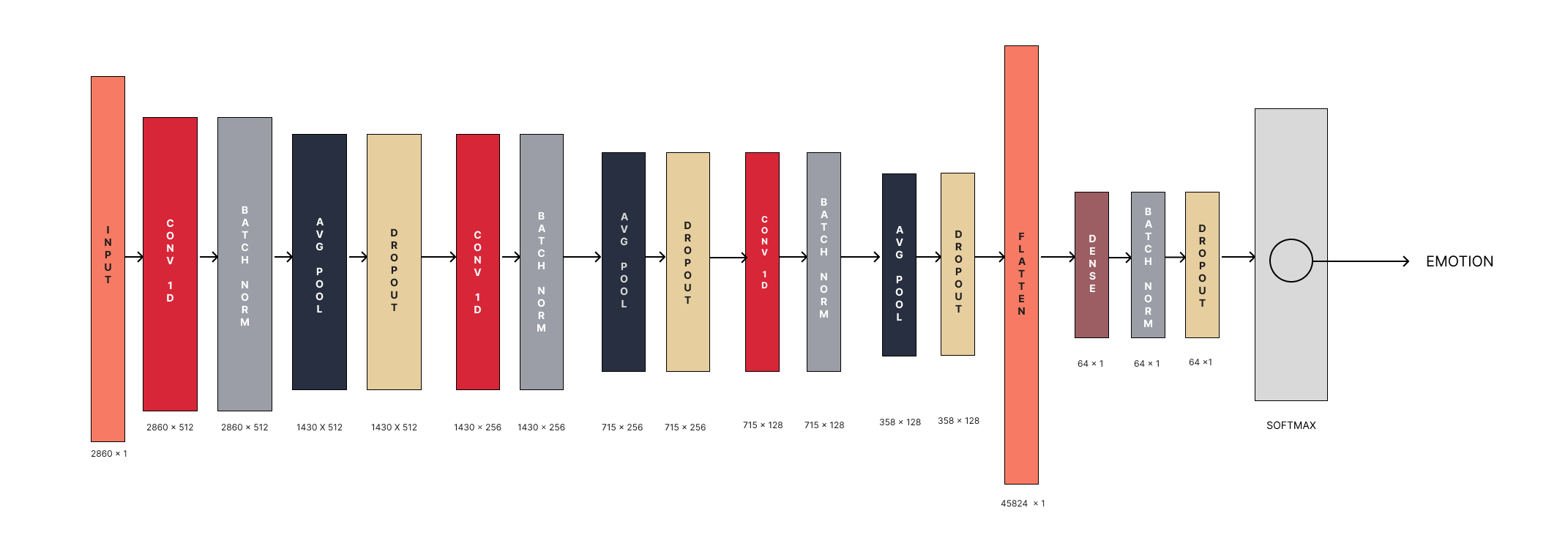}
    \caption{Model architecture for Individual SEGAA}
    \label{fig:segaa}
\end{figure}

\subsubsection{Multi-output Models}
\paragraph{Multi-output MLP:}

    The devised multi-layer perceptron architecture integrates fully connected layers. The input layer is tailored to accommodate the extracted features, leading to subsequent hidden layers, each with 2048, 1024, 512, and 64 neurons as described in figure 4, and employing Rectified Linear Unit (ReLU) activation functions. A dropout layer with a rate of 0.25 is introduced to alleviate overfitting. For label-specific predictions, output layers employ distinct activation functions: sigmoid for gender prediction and softmax for both emotion and age prediction, reflecting the multi-class nature of these tasks. A stochastic Gradient Descent (SGD) optimizer is adopted, featuring a learning rate of 0.0005, a decay rate of 1e-6, momentum set to 0.9, and Nesterov momentum. The model is trained over 200 epochs with a batch size of 32 while monitoring accuracy metrics for gender, emotion, and age predictions on both training and validation datasets. 

\begin{figure}[H]
    \centering
    \includegraphics[width=\linewidth]{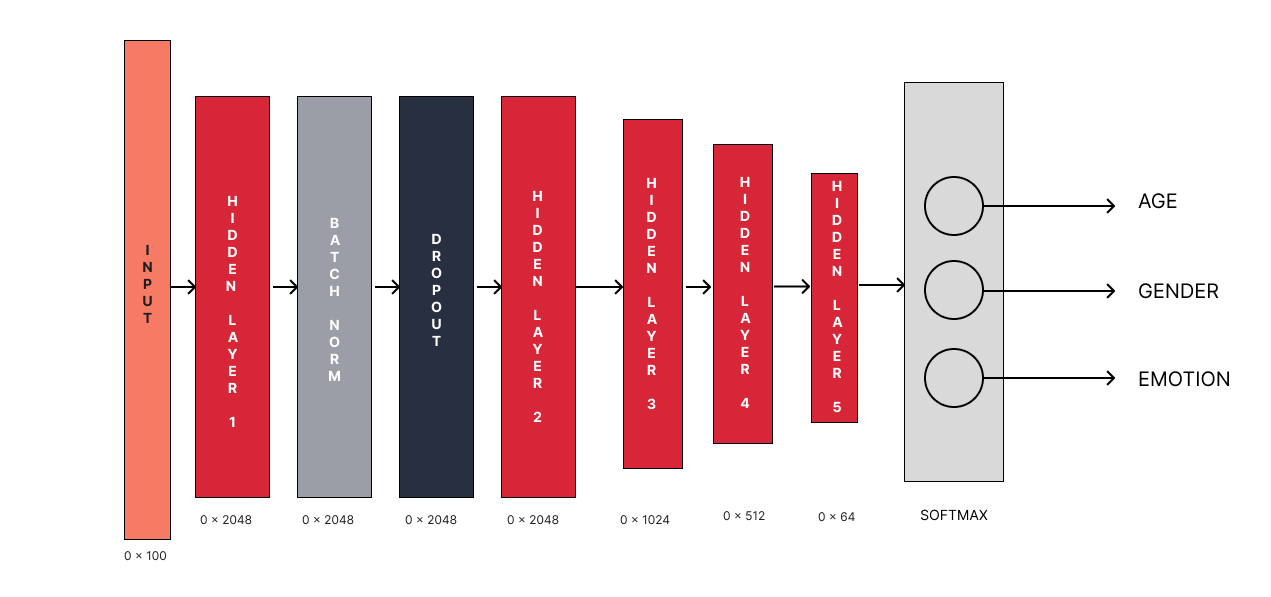}
    \caption{Model architecture for Multi-output MLP}
    \label{fig:multi-mlp}
\end{figure}

\paragraph{Multi-output SEGAA (Gen - 0):}

    In this architecture, the following layers are used: first, a convolutional layer with 256 filters, kernel size of 5, and stride of 1, augmented with batch normalization and then max pooling with pool size 5 and stride 2. Subsequently, a convolutional layer with 128 filters, kernel size of 5, and stride of 1, combined with batch normalization, max pooling with pool size 5 and stride 2, and drop out at a 20\% rate. Lastly, a convolutional layer with 64 filters, kernel size of 5, and stride of 1, along with batch normalization and max pooling, completes the cascade.

    Following these layers, a flatten operation leads to a shared densely connected layer composed of 32 neurons, each enhanced with batch normalization and dropout set at a rate of 20\%. Three distinct output layers then come into play, with softmax activation applied for 'emotion' and 'age' predictions and binary softmax activation for 'gender' predictions. The hyperparameters governing the model's effectiveness were diligently selected: a convolutional kernel size of 5, dropout rate of 20\%, utilization of the Adam optimizer for efficient optimization, and an early stopping strategy with a patience of 5 epochs. A learning rate reduction strategy was also embedded, with patience of 3 epochs and a reduction factor of 0.5.

\paragraph{Multi-output SEGAA:}

    The architecture consists of three convolutional blocks, each followed by Batch Normalization, Max Pooling, and Dropout layers, promoting feature extraction and dimensionality reduction. Subsequently, a flattened layer aggregates the information, leading to a densely connected layer with 64 neurons, further normalized and regularized with Batch Normalization and Dropout as described in figure 5. Three separate output layers cater to each label category: 'emotion,' 'age,' and 'gender,' utilizing softmax activation for the former two and binary softmax for the latter. 
    
    Hyperparameters include a kernel size of 3 and stride of 1 for the Convolutional layers, dropout rates of 0.3, and a Nadam optimizer for optimization. The model is trained over 200 epochs using a batch size of 16, monitored by validation accuracy metrics. Early stopping and learning rate reduction callbacks are utilized to ensure optimal convergence. 

\begin{figure}[H]
    \centering
    \includegraphics[width=\linewidth]{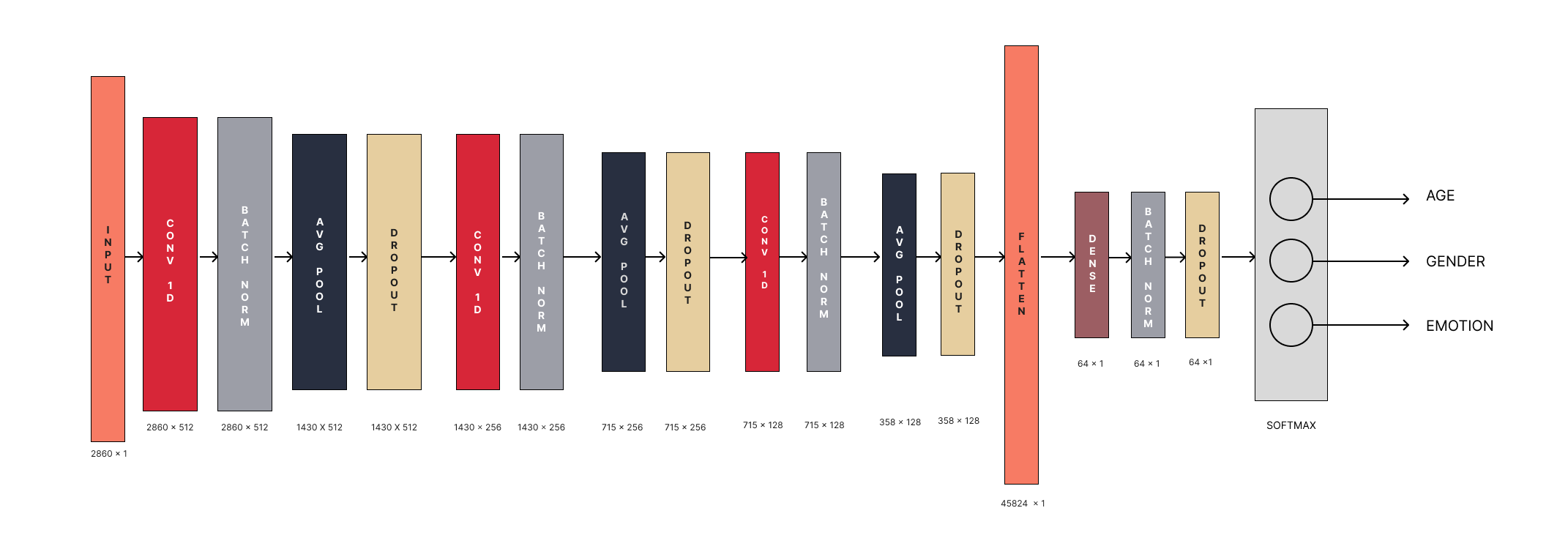}
    \caption{Model architecture for Multi-output SEGAA (Speech-based Emotion, Gender, Age Analysis)}
    \label{fig:multi-segaa}
\end{figure}

\subsubsection{Sequential Models}

    In this experimental setup, the individual models described in Section 3.3.1 were cascaded sequentially to predict the first variable from the speech features, followed by predicting the second variable from the speech features and the previously predicted first variable, and finally predicting the third variable from the speech features and the previously predicted second variable as described in figure 5. This experiment was conducted in three different sequences: first for the variables emotion, gender, and age; then for gender, age, and emotion; and lastly for age, emotion, and gender.

\begin{figure}[H]
    \centering
    \includegraphics[width=\linewidth]{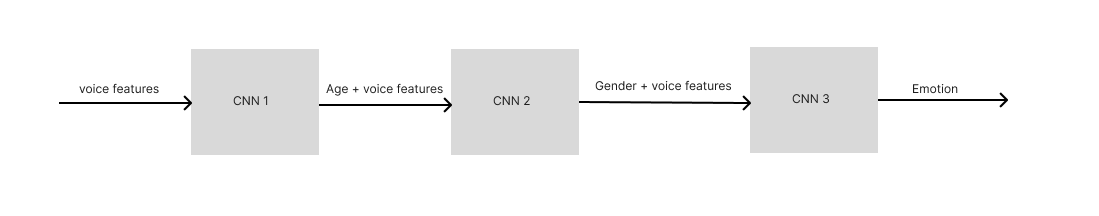}
    \caption{Model architecture for Sequential models}
    \label{fig:seq-model}
\end{figure}

\section{Results}
\label{sec:results}

    This research investigates the task of speech-based detection of emotion, gender, and age using a diverse range of machine-learning models. Our primary objective is to develop accurate and robust models for classifying these three attributes from audio data. The experimental framework includes individual and multi-output Multi-Layer Perceptron (MLP) as well as SEGAA architectures. Additionally, we investigate unique sequences for predicting these attributes with our SEGAA models. 

\subsection{Individual Models}
\subsubsection{Emotion Detection}

    For emotion detection, both individual SEGAA and MLP models exhibit robust performance. The SEGAA model attains an accuracy of 96\%, while the MLP model achieves 94\% accuracy. These models demonstrate well-balanced precision, recall, and F1 scores, ranging from 0.94 to 0.96.

\subsubsection{Gender Detection}

    The individual models demonstrate exceptional accuracy for gender detection. The SEGAA model attains a flawless accuracy of 100\%, while the MLP model achieves 98\%. Precision, recall, and F1 scores consistently reach 1.00 for both models.

\subsubsection{Age Detection}

Age detection, too, benefits from the individual SEGAA and MLP models, with the SEGAA model achieving 95\% accuracy and the MLP model reaching 92\%. These models display high precision, recall, and F1 scores, ranging from 0.92 to 0.95.

\subsection{Multi-output Models}

Our investigation extends to multi-output SEGAA and MLP models, which tries to simultaneously predict emotion, gender, and age in that order. These models yield competitive results with those of the individual models, with accuracy values spanning from 84\% to 99\% across generations. Although the multi-output SEGAA Gen-0 model maintains strong precision and F1 scores, there is a slight decrease in recall for certain attributes, particularly in emotion detection. In contrast, the proposed SEGAA model effectively addresses these limitations, demonstrating exceptional performance across all aspects, encompassing emotion, age, and gender detection, which is highlighted in Figure 7.

\subsection{Sequential Models}

    Exploration of various sequences for predicting emotion, gender, and age using SEGAA models reveals distinct strengths across different orderings. The EGA sequence (Emotion → Gender → Age) notably excels in gender detection, delivering respectable accuracy in emotion detection. This sequence demonstrated has consistent F1 scores above 0.90 for all three attributes which talks extensively about the sequential model metrics.\\

\begin{table}[H]
\parbox{.5\linewidth}{
    \caption{MLP \& SEGAA Model Metrics}
    \label{tab:my-table-1}
    \begin{tabular}{@{}lllll@{}}
    \toprule
    Target           & Accuracy      & Precision     & Recall        & F1 Score      \\ \midrule
    \multicolumn{5}{c}{Individual MLP}                                               \\ \midrule
    Age              & 92\%          & 0.92          & 0.92          & 0.92          \\
    Emotion          & 94\%          & 0.94          & 0.94          & 0.94          \\
    Gender           & 98\%          & 0.98          & 0.98          & 0.98          \\ \midrule
    \multicolumn{5}{c}{Multi-output MLP}                                             \\ \midrule
    Age              & 94\%          & 0.94          & 0.94          & 0.94          \\
    Emotion          & 95\%          & 0.95          & 0.95          & 0.95          \\
    Gender           & 98\%          & 0.98          & 0.98          & 0.98          \\ \midrule
    \multicolumn{5}{c}{Individual SEGAA}                                             \\ \midrule
    Age              & 95\%          & 0.95          & 0.95          & 0.95          \\
    Emotion          & 96\%          & 0.96          & 0.96          & 0.96          \\
    Gender           & 100\%         & 1             & 1             & 1             \\ \midrule
    \multicolumn{5}{c}{Multi-output SEGAA (Gen-0)}                                   \\ \midrule
    Age              & 90\%          & 0.93          & 0.9           & 0.9           \\
    Emotion          & 84\%          & 0.86          & 0.84          & 0.84          \\
    Gender           & 98\%          & 0.98          & 0.98          & 0.98          \\ \midrule
    \multicolumn{5}{c}{\textbf{Multi-output SEGAA}}                                  \\ \midrule
    \textbf{Age}     & \textbf{94\%} & \textbf{0.96} & \textbf{0.94} & \textbf{0.95} \\
    \textbf{Emotion} & \textbf{95\%} & \textbf{0.95} & \textbf{0.95} & \textbf{0.95} \\
    \textbf{Gender}  & \textbf{99\%} & \textbf{0.99} & \textbf{0.99} & \textbf{0.99} \\ \bottomrule
    \end{tabular}
}
\parbox{.5\linewidth}{
        \caption{Sequential Model Metrics}
        \label{tab:my-table-2}
        \begin{tabular}{@{}lllll@{}}
        \toprule
        Target     & Accuracy   & Precision   & Recall   & F1 Score   \\ \midrule
        \multicolumn{5}{c}{CNN Sequence 1 (Gender → Age   → Emotion)} \\ \midrule
        Gender     & 99\%       & 0.99        & 0.99     & 0.99       \\
        Age        & 94\%       & 0.95        & 0.95     & 0.95       \\
        Emotion    & 94\%       & 0.94        & 0.94     & 0.94       \\ \midrule
        \multicolumn{5}{c}{CNN Sequence 2 (Age → Emotion   → Gender)} \\ \midrule
        Age        & 94\%       & 0.94        & 0.94     & 0.94       \\
        Emotion    & 91\%       & 0.91        & 0.91     & 0.91       \\
        Gender     & 99\%       & 0.99        & 0.99     & 0.99       \\ \midrule
        \multicolumn{5}{c}{CNN Sequence 3 (Emotion →   Gender → Age)} \\ \midrule
        Emotion    & 95\%       & 0.95        & 0.95     & 0.95       \\
        Gender     & 99\%       & 0.99        & 0.99     & 0.99       \\
        Age        & 92\%       & 0.93        & 0.92     & 0.92       \\ \midrule
        \multicolumn{5}{c}{MLP Sequence (Emotion →   Gender → Age)}   \\ \midrule
        Emotion    & 91\%       & 0.91        & 0.91     & 0.91       \\
        Gender     & 90\%       & 0.91        & 0.9      & 0.91       \\
        Age        & 92\%       & 0.93        & 0.92     & 0.92       \\ \bottomrule
        \end{tabular}
        
}

\end{table}

    Table 1 encapsulates the performance metrics of individual and multi-output models, providing a detailed assessment of accuracy, precision, recall, and F1 scores for age, emotion, and gender detection. For individual models, both SEGAA and MLP models showcase high accuracy, with the SEGAA model achieving 96\% accuracy in emotion detection and demonstrating gender detection at 100\%, with more information on the models highlighted in Figure 6. Notably, the multi-output models, particularly the SEGAA model, exhibit competitive results across all attributes, achieving a remarkable 99\% accuracy in gender detection. While the Gen-0 displays slightly lower scores in emotion detection, SEGAA addresses this limitation effectively as the Nadam optimizer present in SEGAA uses Nesterov’s Accelerated Gradient as an extension to the Adam optimizer used in Gen-0. The multi-output MLP model also maintains high accuracy and balanced metrics, emphasizing the performance of this model category in simultaneous prediction tasks.

    While all models consistently deliver competitive performance across accuracy and other metrics, it is noteworthy that the proposed model (Multi-output SEGAA) emerges as the most efficient choice. These models offer commendable accuracy while preserving efficiency, an important consideration in real-world applications where computational resources and low latency are necessary.
    
\begin{figure}[H]
\centering
\begin{minipage}{.5\textwidth}
  \centering
  \begin{subfigure}[t]{\linewidth}
  \includegraphics[width=\linewidth]{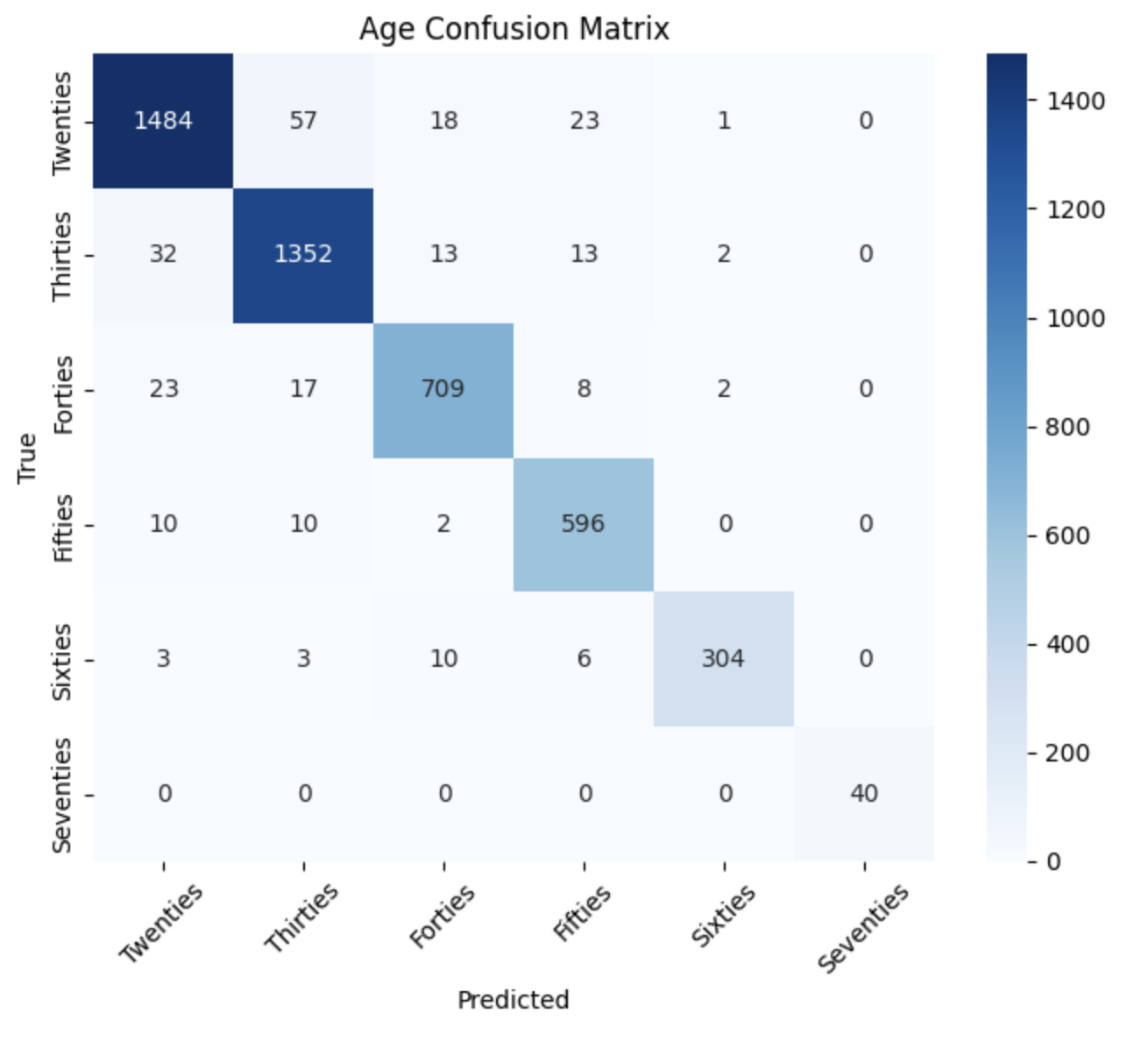}
  \label{fig:test1}
  \caption{Age Confusion Matrix}
  \end{subfigure}
\end{minipage}%
\begin{minipage}{.5\textwidth}
  \centering
  \begin{subfigure}[t]{\linewidth}
  \includegraphics[width=\linewidth]{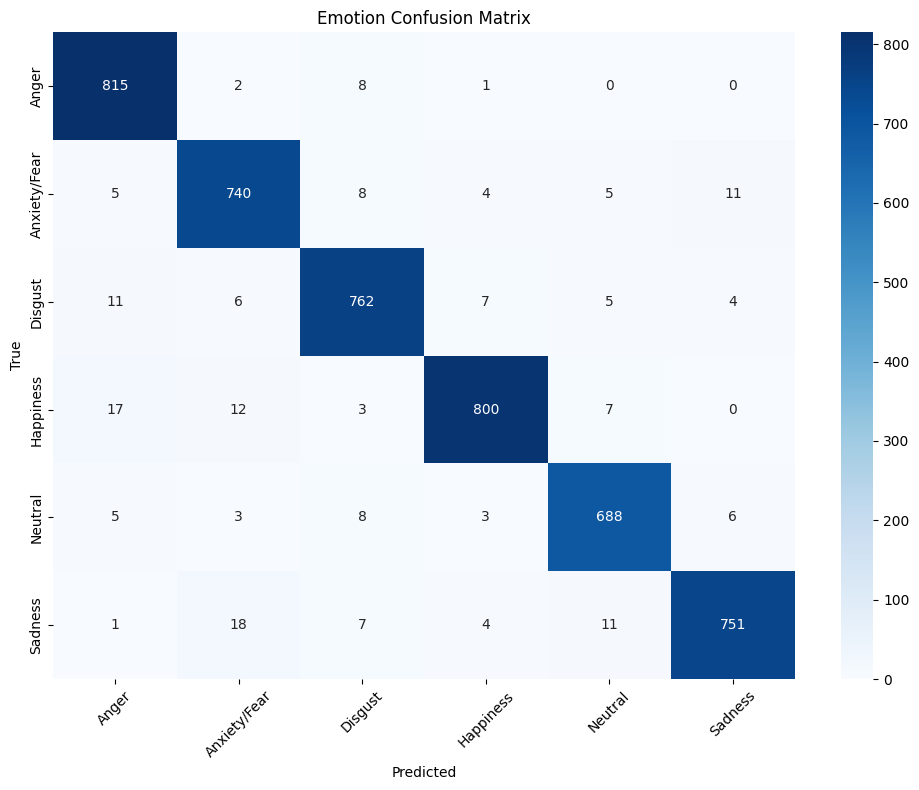}
  \label{fig:test2}
  \caption{Emotion Confusion Matrix}
  \end{subfigure}
\end{minipage}
  \centering
  \begin{subfigure}[t]{0.5\linewidth}
  \includegraphics[width=\linewidth]{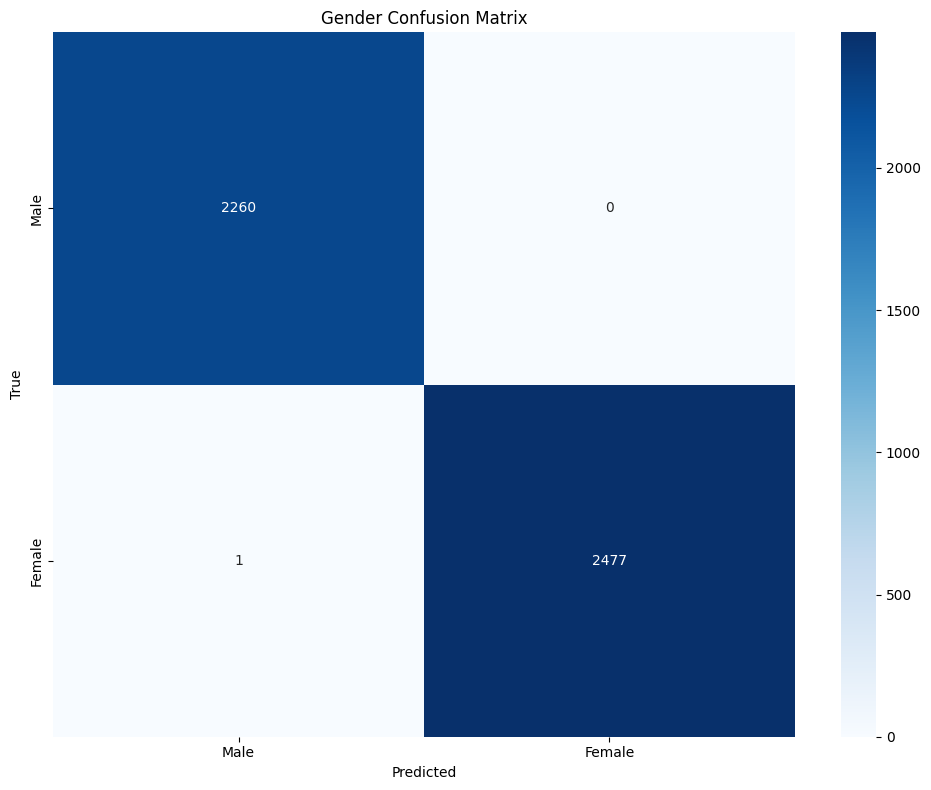}
  \label{fig:test3}
  \caption{Gender Confusion Matrix}
  \end{subfigure}
\caption{Confusion Matrices for Individual SEGAA}
\end{figure}

\FloatBarrier
    Predicting gender through audio has been a pretty easy task for the individual models. This can be seen in Figure 7a and 7b, where there are barely any misclassifications, and a high number of true positives for each of the categories. The same can be said for the task of emotion prediction for these models, as there aren’t many misclassifications, compared to the number of right predictions. However, they struggle a bit more when it comes to identifying "neutral" and "anxiety/fear" emotions. The models also excel at predicting the age accurately, but it's important to note that the dataset is imbalanced, as there are fewer examples of people in their "sixties" and "seventies" in the dataset compared to other age groups.

\begin{figure}[H]
\centering
\begin{minipage}{.5\textwidth}
  \centering
  \begin{subfigure}[t]{\linewidth}
  \includegraphics[width=\linewidth]{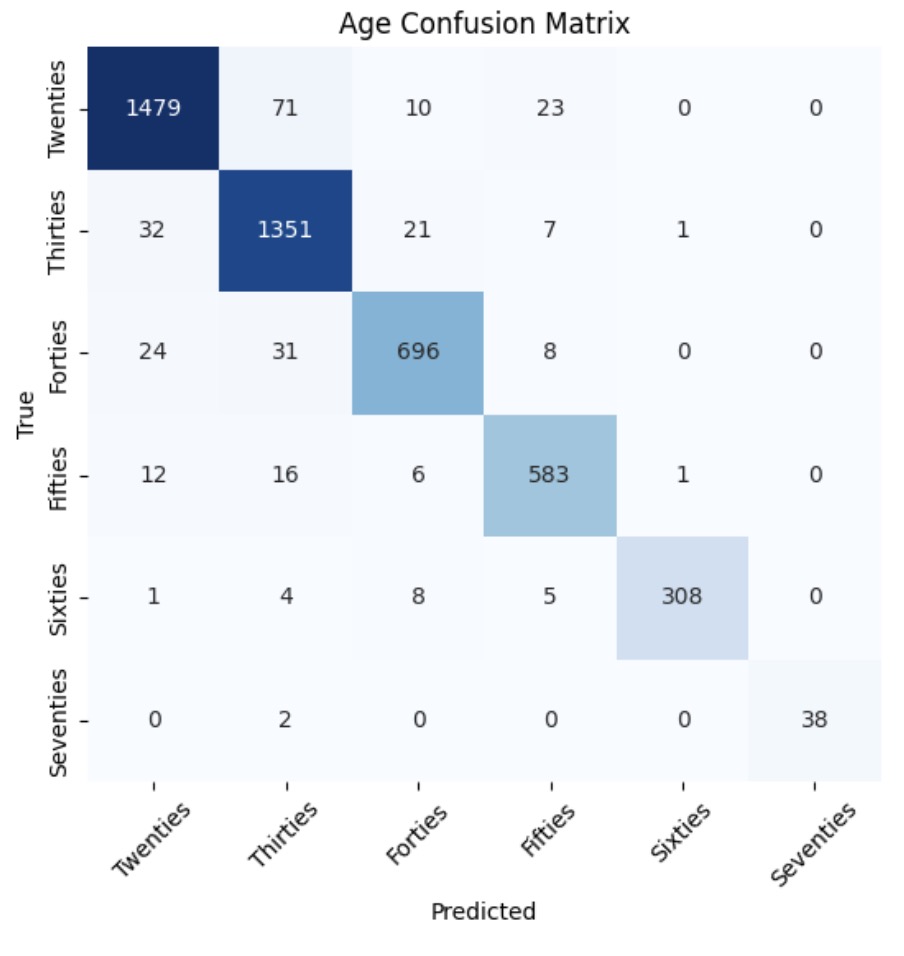}
  \label{fig:test4}
  \caption{Age Confusion Matrix}
  \end{subfigure}
\end{minipage}%
\begin{minipage}{.5\textwidth}
  \centering
  \begin{subfigure}[t]{\linewidth}
  \includegraphics[width=\linewidth]{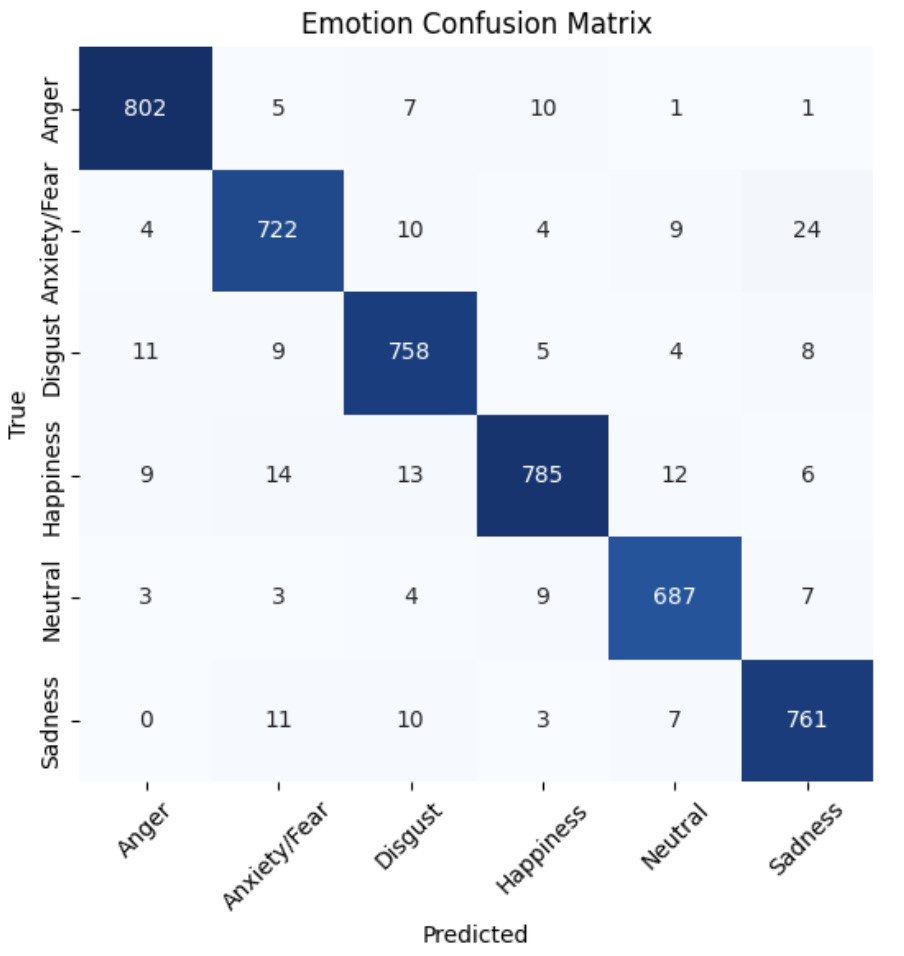}
  \label{fig:test5}
  \caption{Emotion Confusion Matrix}
  \end{subfigure}
\end{minipage}
  \centering
  \begin{subfigure}[t]{0.5\linewidth}
  \includegraphics[width=\linewidth]{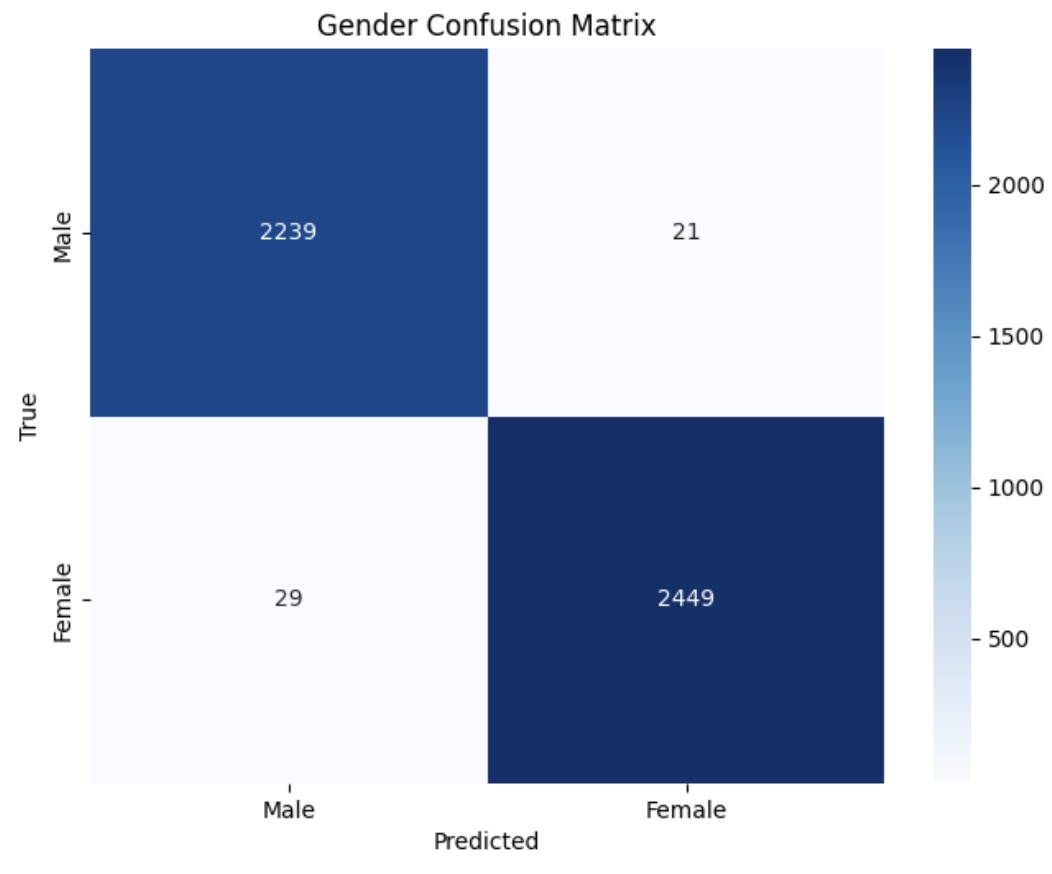}
  \label{fig:test6}
  \caption{Gender Confusion Matrix}
  \end{subfigure}
\caption{Confusion Matrices for Multi-output SEGAA}
\end{figure}

\FloatBarrier
    Comparing the results for Individual SEGAA and Multi-output SEGAA shows that the model itself is stable, since there is relatively little to no change in the classifications made by the two models. The Multi-output SEGAA is comparable to the Individual SEGAA in terms of performance. Individual SEGAA outperforms the Multi-output SEGAA models marginally in all the tasks, as shown in Figure 8a and 8b, but the training time taken by the multi-output SEGAA outweighs this merit, as they are already extremely close in performance.

\FloatBarrier

\begin{figure}[H]
\centering
\begin{minipage}{.5\textwidth}
  \centering
  \begin{subfigure}[t]{\linewidth}
  \includegraphics[width=\linewidth]{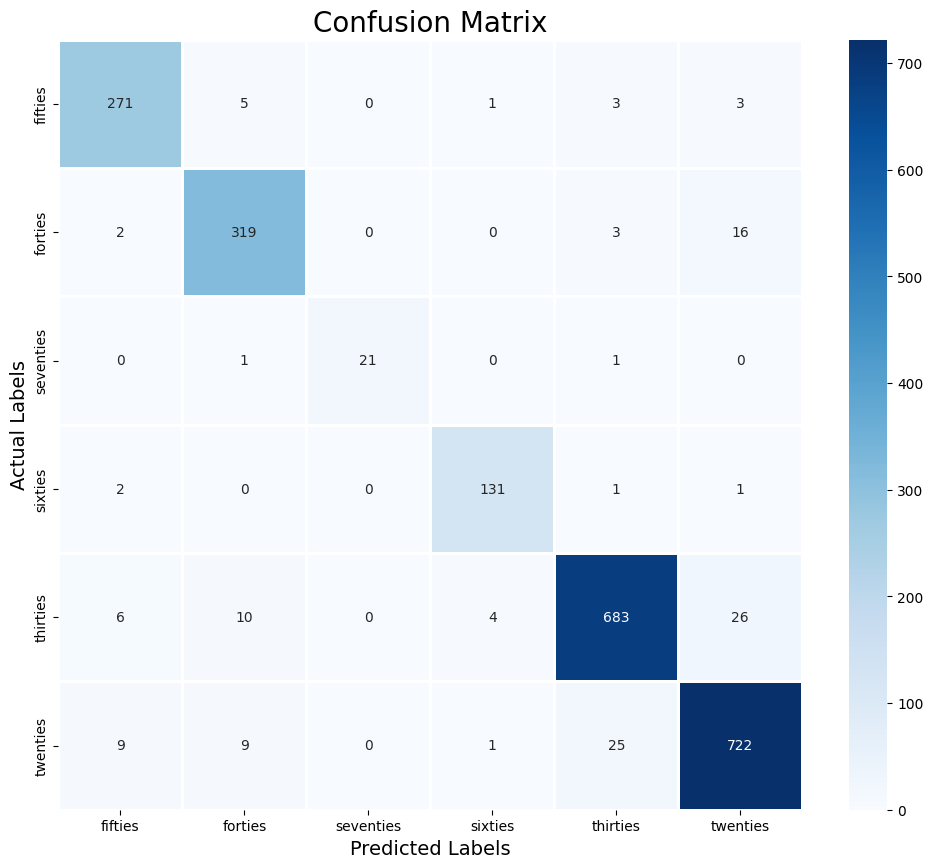}
  \label{fig:test7}
  \caption{Age Confusion Matrix}
  \end{subfigure}
\end{minipage}%
\begin{minipage}{.5\textwidth}
  \centering
  \begin{subfigure}[t]{\linewidth}
  \includegraphics[width=\linewidth]{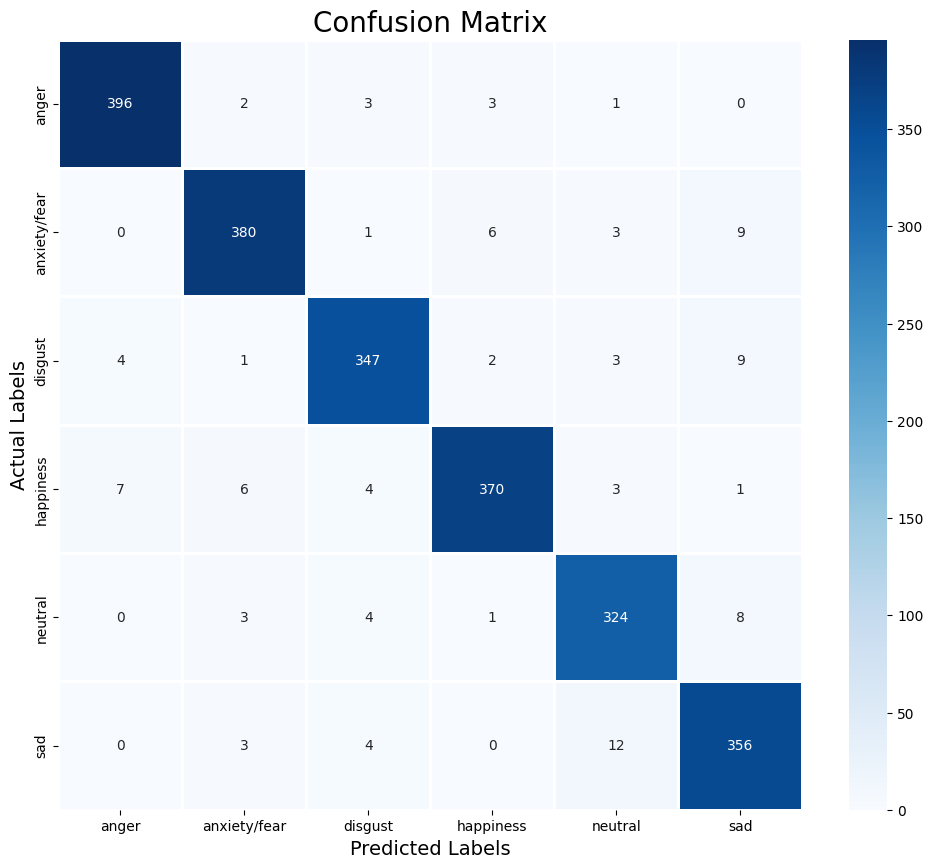}
  \label{fig:test8}
  \caption{Emotion Confusion Matrix}
  \end{subfigure}
\end{minipage}
  \centering
  \begin{subfigure}[t]{0.5\linewidth}
  \includegraphics[width=\linewidth]{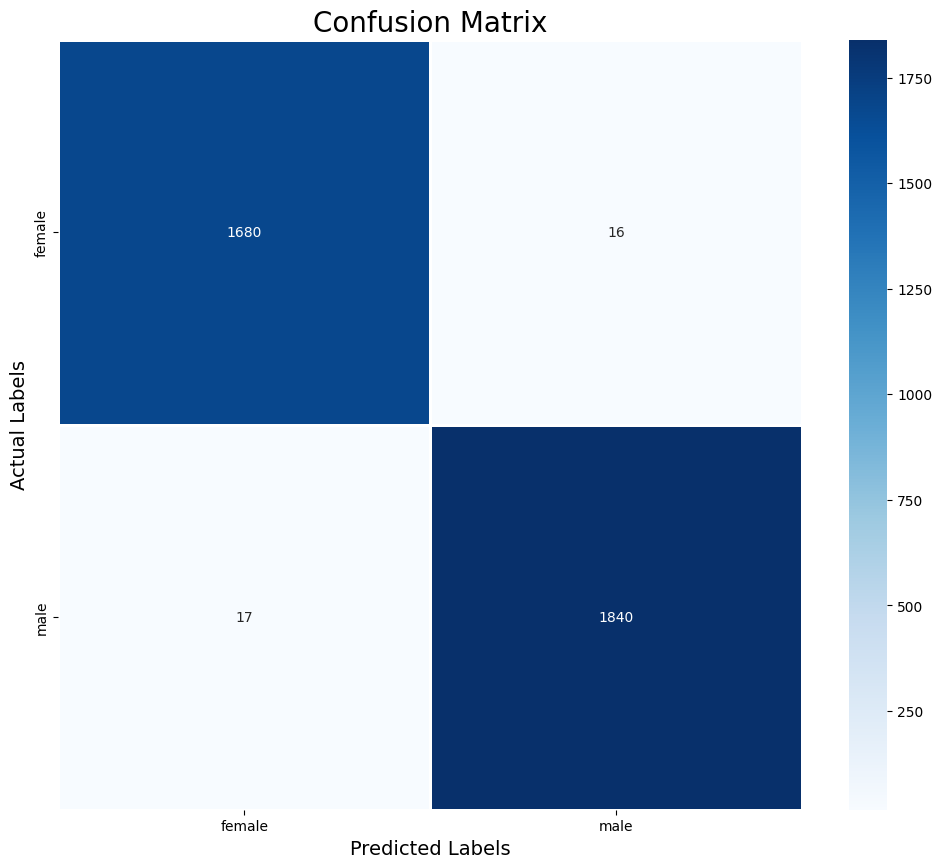}
  \label{fig:test9}
  \caption{Gender Confusion Matrix}
  \end{subfigure}
\caption{Confusion matrices for Sequence 3: (Emotion → Gender → Age)}
\end{figure}

    The performance of Emotion → Gender → Age sequence showcased in Figure 9a and 9b is better than the rest, but it still isn't as good as the individual or multi-output SEGAA models, which is because of the sum of errors due to error from one model propagating to the next, which adds up. This is apparent in the confusion matrices, where the model hasn't been able to classify the age of seventies as well as the SEGAA model, along with the other classes.

\section{Conclusion}

    In our experimentation, we conducted a rigorous comparative assessment of univariate and multi-output models to predict gender, age, and emotion from speech data, a crucial task with applications spanning various domains. Our analysis unveiled a noteworthy phenomenon associated with sequentially chaining univariate models: this approach increased error propagation as the inaccuracies generated by preceding models were amplified in subsequent stages. Despite this, univariate models exhibited slightly superior accuracy and F1 scores performance compared to multi-output models.

    It is essential to emphasize that our experimental findings suggest that SEGAA demonstrates a level of predictive capability comparable to univariate models. Notably, it excels in capturing the intricate interrelationships between the variables and speech inputs with commendable efficiency. Furthermore, SEGAA achieves this without compromising runtime efficiency, making them an attractive alternative for addressing the complexities of predicting gender, age, and emotion from speech data. These insights, we believe, hold valuable implications for both researchers and practitioners in the field.

\bibliographystyle{unsrtnat}
\bibliography{refs}
\end{document}